\documentclass[prd,aps,twocolumn,preprintnumbers,showpacs,nofootinbib,superscriptaddress,notitlepage]{revtex4-2}
\usepackage{hyperref}
\usepackage{comment}
\usepackage{bm}
\usepackage{amsmath}
\usepackage{slashed}
\usepackage{xcolor}
\usepackage{graphicx}
\usepackage{caption}
\usepackage{subcaption}
\usepackage[normalem]{ulem}
\usepackage[utf8]{inputenc}
\usepackage[T1]{fontenc} 

\newcommand{\beq}{\begin{eqnarray}}
\newcommand{\eeq}{\end{eqnarray}}

\begin{document}

\title{Large-Momentum Effective Theory vs. Short-Distance Operator Expansion \\ Contrast and Complementarity}

\author{Xiangdong Ji}
\email{xji@umd.edu}
\affiliation{Maryland Center for Fundamental Physics,
Department of Physics, University of Maryland,
College Park, Maryland 20742, USA}

\date{\today{}}
\begin{abstract}

Although equivalent in the infinite-momentum limit, large-momentum effective theory (LaMET) and short-distance operator product expansion (SD-OPE) are two different approaches to extract parton distribution functions (PDFs) from coordinate-space correlation functions in large-momentum hadrons. LaMET implements a momentum-space expansion in $\Lambda_{\rm QCD}/[x(1-x)P^z]$ to directly calculate PDFs $f(x)$ in a middle region of Bjorken $x\in [x_{\rm min}\sim \Lambda_{\rm QCD}/P^z, x_{\rm max}\sim 1-x_{\min}]$. SD-OPE applies perturbative QCD at small Euclidean distances $z$ to extract a range $[0,\lambda_{\rm max}]$ of leading-twist correlations, $h(\lambda=zP^z)$, corresponding to the Fourier transformation of PDFs. Similar to the quantum mechanical uncertainty principle, an incomplete leading-twist correlation cannot be readily converted to a momentum-space local distribution, and the methods to solve the ``inverse problem'' involve essentially modelling of the missing information beyond $\lambda_{\rm max}$. On the other hand, short-distance correlations, along with the expected end-point asymptotics, can be used to phenomenologically fit the PDFs in the LaMET-complementary regions: $x\in [0,x_{\rm min}]$ and $[x_{\rm max}, 1]$. We use the recent results of the pion valence quark distribution from the ANL/BNL collaboration to demonstrate this point. 
\end{abstract} \maketitle

\section{Introduction}

Parton distribution functions (PDFs) are one of the most important properties of the proton
and neutron, and they are necessary, among others,
for predicting high-energy cross sections at the Large Hadron Collider. In the past, the best knowledge of the PDFs has been gleaned from fitting parametrizations to the data obtained from high-energy experiments over several decades~\cite{Gao:2017yyd}.
In recent years, calculating parton distributions from the first principles of quantum chromodynamics (QCD) has gained considerable momentum among 
lattice-QCD practioners. Much progress has been made and some recent summaries can be found in Refs.\cite{Lin:2017snn,Cichy:2018mum,Ji:2020ect,Constantinou:2020hdm}. 

For some time, first-principles studies of PDFs are limited to their moments which correspond to the matrix elements of local twist-2 operators~\cite{Martinelli:1987bh,Gockeler:1995wg,Lin:2017snn}. Since the 1990's, methods have been proposed to directly calculate the correlation functions of operators/currents whose expansion in the large momentum transfer $q$ or Euclidean short distance $z$ (SD-OPE) is dominated by towers of twist-2 operators, thus effectively providing a method to calculate a few lower moments at once~\cite{Aglietti:1998ur,Detmold:2005gg,Braun:2007wv,Radyushkin:2017cyf,Chambers:2017dov,Ma:2017pxb,Bali:2018spj,Detmold:2021uru}. The correlations with a finite range in $z$, although insufficient to determine the $x$-dependent PDFs from first principles, have also been used to phenomenologically constrain PDFs using various inverse-problem methods, including fitting parameterized functions as in the global analyses of PDFs ~\cite{Karpie:2019eiq,Sufian:2019bol,Joo:2019jct,Joo:2019bzr,Sufian:2020vzb,Bhat:2020ktg,Gao:2020ito,DelDebbio:2020rgv,Joo:2020spy,Karpie:2021pap}. 

An alternative method to calculate $x$-dependent PDFs on lattice has been proposed following Feynman's original idea about partons~\cite{Bjorken:1969ja, Feynman:1969ej}, according to which, they are the effective constituents of hadrons when the latter travel at the speed of light. Thus PDFs are momentum densities of quarks and quarks in the infinite
momentum frame with the limit
taken, in field theories, before regularizing the ultraviolet (UV) divergences. 
In actual calculations, one can approximate PDFs by 
the momentum distributions of quarks and gluons at finite hadron momentum $P^z$, which are calculated with UV cut-off much larger than $P^z$~\cite{Ji:2013dva}. Different UV limits can be matched by QCD perturbation theory using effective field theory methods~\cite{Ji:2014gla}, thanks to asymptotic freedom. The 
resulting power corrections in $(\Lambda_{\rm QCD}/P^z)^2$, where $\Lambda_{\rm QCD}$ is the strong interaction scale, can in principle be systematically studied~\cite{Chen:2016utp}. This method has been named large momentum effective theory or LaMET, which is a general framework to calculate parton physics and light-cone correlations far beyond the scope of the collinear PDFs, including generalized parton distributions (GPDs) and transverse momentum dependent (TMD) PDFs, as well as light-front wave functions~\cite{Ji:2020ect,Ji:2021znw}.

When limited to collinear PDFs, SD-OPE and the framework of LaMET are equivalent in the infinite momentum limit~\cite{Izubuchi:2018srq}. However, they are quite different at finite $P^z$ where the actual lattice calculations are performed~\cite{Ji:2020byp, Ji:2020brr}. SD-OPE extracts the leading-twist short-range correlations in a range of light-cone distance $[0,\lambda_{\rm max}\sim (0.2~{\rm fm})P^z ]$, whereas LaMET 
calculations yield twist-2 $x$-space PDFs in a range of momentum fraction $x \in [x_{\rm min}\sim \Lambda_{\rm QCD}/P^z, x_{\rm max}\sim 1-x_{\rm min}]$, where
the kinematic limits ($\lambda_{\rm max}=\infty, x_{\rm min}=0, x_{\rm max}=1$) are reached only at $P^z=\infty$. The short-distance correlations encode global information 
about PDFs, but cannot readily determine their local properties in $x$-space, whereas LaMET produces directly local PDFs at given $x$'s. Since the high-energy experiments measure momenta of particles, the LaMET expansion appears to be a more natural approach to calculating, rather than fitting, PDFs. 

In this article, we discuss contrast and complementarity of these two widely-studied approaches to collinear PDFs. In particular, we study how to re-use the coordinate-space data to phenomenologically determine the PDFs outside the momentum expansion region. For a given large $P^z$, LaMET analysis of lattice correlation data produces the most accurate information on PDFs in a range of $x$ where systematics are under control~\cite{Ji:2020byp,Gao:2021dbh}. In the end-point regions $[0,x_{\rm min}]$ and $[1-x_{\rm max},1]$, however, the expansion breaks down. However, one may be able to constrain PDFs in these regions using  
\begin{itemize}
    \item The end-point behavior known from theory consideration and phenomenology. We know, e.g.,  light-cone 
    PDFs must vanish at $x=1$. Moreover, small-$x$ physics is constrained by Regge behavior~\cite{Ioffe:1985ep} and large-$x$ behavior by perturbative QCD power counting~\cite{Brodsky:1973kr}. 
    \item Low moments of PDFs or short-distance correlations, which control the global properties of PDFs. 
\end{itemize}
Of course, this information has already been used in the analysis of lattice data in the SD-OPE approach~\cite{Karpie:2019eiq,Sufian:2019bol,Joo:2019jct,Joo:2019bzr,Sufian:2020vzb,Bhat:2020ktg,Gao:2020ito,DelDebbio:2020rgv,Joo:2020spy,Karpie:2021pap}. However, we suggest to better use it in a different context, namely, to phenomenologically bridge the gaps in LaMET analyses. In fact, one might even want to argue that this is the most appropriate use of the 
phenomenological parametrization and short-distance correlation data. From the global analysis perspective, the LaMET results provide the key constraints on PDFs in the middle $x$ region, which must be satisfied in any fitting.

We will begin by reviewing some of the most salient features of SD-OPE and LaMET in the next two sections. We then contrast and complement the two approaches in Sec. IV. Throughout the discussions, we will use the pion valence PDF calculations by the ANL/BNL group as illustration~\cite{Gao:2020ito,Gao:2021hxl,Gao:2021dbh}. The strategy is also applicable to other PDF calculations, such as those for GPDs and distribution amplitudes (DAs), and even TMDs. 

\section{Moments, Short-distance Correlations and 
Global Properties of Parton Distributions}

Lattice QCD calculations in parton physics started from the matrix elements of local twist-2 operators,
which are PDF moments~\cite{Martinelli:1987bh,Gockeler:1995wg}, e.g.,
\begin{equation}
    M_n(\mu) = \int^1_{0} \Big[f_q (x,\mu) +(-1)^n f_{\bar q}(x,\mu)\Big]x^{n-1} dx \ , 
\end{equation}
where $f_{q,\bar q}(x)$ are unpolarized quark and anti-quark distributions of flavor $q$. 
$\mu$ indicates renormalization scheme and scale, chosen usually in dimensional regularization and modified minimal subtraction ($\overline{\rm MS}$).

Moments are meant to capture {\it global properties} of a distribution. For instance, when $n=1$, the moment
gives the area under the curve $f_q(x)-f_{\bar q}(x)$, which counts the total number of
valence quarks. For PDFs, the first moments sometime provide a strong constraint on the low-$x$ behavior where they tend to rise due to the presence of a large number of sea quarks and gluons. In fact, when measuring the first moment of the proton's $g_1(x)$ spin structure function, the small $x$ contribution has a large uncertainty due to the unknown small-$x$ behavior~\cite{E143:1994vcg}. There is also a significant uncertainty for the total gluon helicity $\Delta G$
from the small $x$~\cite{deFlorian:2014yva}. 

Higher-order moments provide additional global information about PDFs. Of course, As $n$ gets larger, $x^n$ weighs more and more toward the $x\sim 1$ region, and the moments become sensitive to the large-$x$ behavior only. For example, if the distribution goes like $(1-x)^\beta$ near $x=1$, the higher-order moments are strongly correlated with the value of $\beta$~\cite{Gao:2020ito}. However, unless one knows many of them, few moments do not give us precise information about the value of $f(x)$ at a particular $x$. 

To translate lower moments into local-$x$ information about PDFs, one usually resorts to models. From general 
physics considerations, the PDFs
vanish at $x=1$ as $(1-x)^\beta$, and
grow like $x^\alpha$ at small $x$~\cite{Ioffe:1985ep}. This suggests a  three-parameter model for the quark distribution 
\begin{equation}
   f_q(x) = Ax^\alpha(1-x)^\beta \ , 
\end{equation}
and similar for the antiquark. This simple model has been widely used to fit parton distributions although it is hard to justify  in the middle-$x$ region. If taking it seriously, one only needs three moments to get a PDF, which in fact has been quite successful in phenomenology~\cite{Detmold:2003tm,Alexandrou:2021mmi}. A more sophisticated model would be to add a factor which is a smooth function of $x$ with additional parameters~\cite{Dulat:2015mca}. 

Lower-order moments are closely related to the short-range behavior of the
twist-2 correlation defined from PDFs, 
\begin{equation}
     h(\lambda,\mu^2) = \int^1_{-1} {dx}
     e^{i\lambda x} f(x, \mu^2) \ . 
\end{equation}
Expanding the right hand side at small $\lambda$ yields, 
\begin{equation}
    h(\lambda,\mu^2) = \sum_{n=0}^\infty M_{n+1}(\mu^2) \frac{(i\lambda)^n}{n!}
    \label{eq:expand}
\end{equation}
where the Taylor-coefficients are just moments.
Therefore, short distance or small-$\lambda$ correlations are mostly determined by the lowest
few moments. 

\begin{figure}[tbh]
\centering
\includegraphics[width=0.47\textwidth]{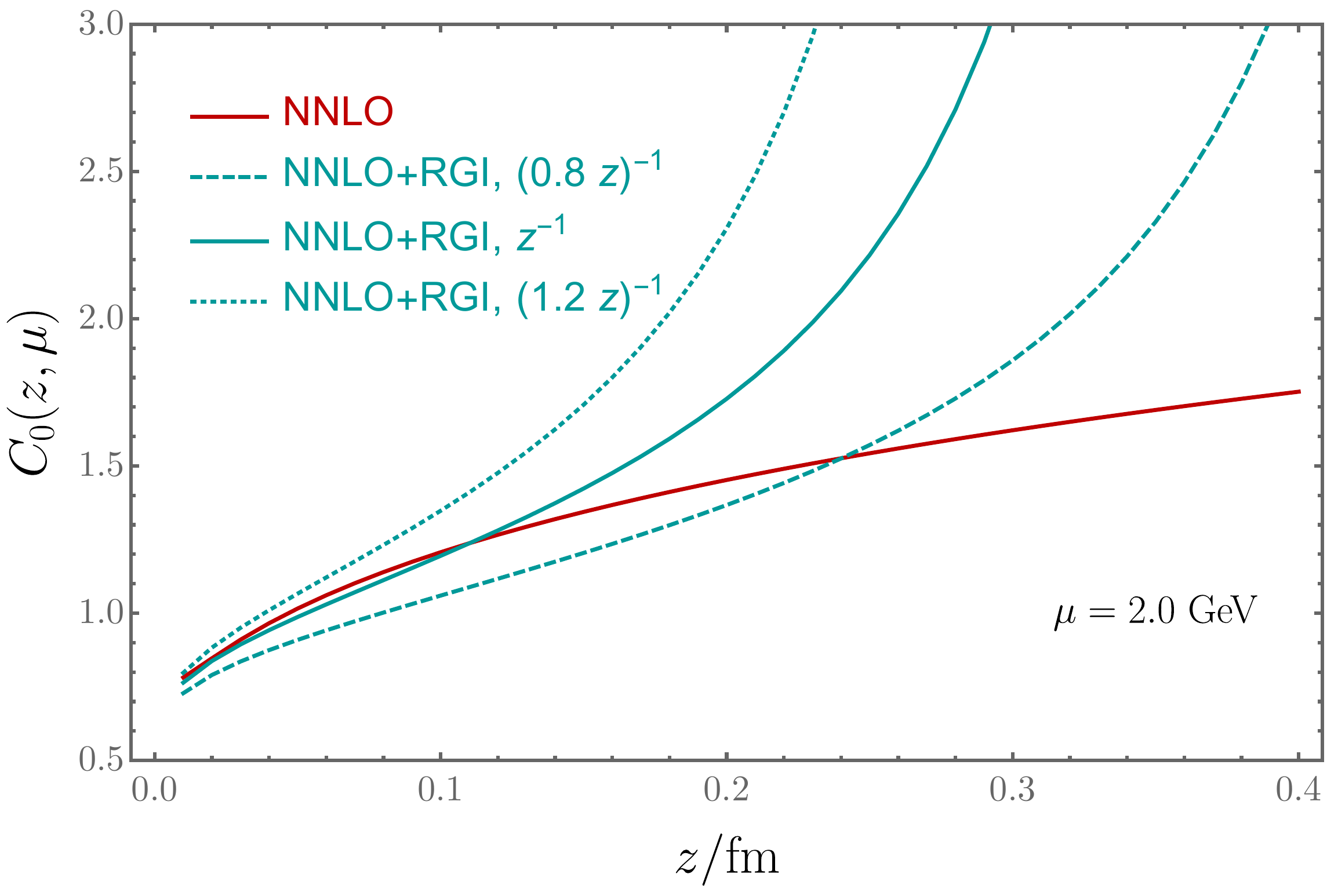}
\centering
\caption{The Wilson coefficient of the number operator as a function of $z$, appearing in the quasi-PDF operator in short-distance expansion~\cite{Gao:2021dbh}. The red line is next-to-next-to leading order result. The resummed results (RGI) are shown as solid lines, which quickly become divergent between $z=0.2$ to $z=0.3$ fm.}
    \label{fig:wilson}
\end{figure}

The SD-PDF approach to calculating PDFs starts from some Euclidean correlators, e.g., $H(\lambda, z^2)=\langle P^z| J(z)J(0)|P^z\rangle$ where, without loss of generality, $J$ is a composite operator of quarks and gluon fields, $z$ is the spatial separation along the $z$ direction, and $\lambda = zP^z$. At small $z^2\ll 1/\Lambda_{\rm QCD}^2$, 
one has the SD-OPE or operator product expansion (OPE)~\cite{Braun:2007wv,Ma:2017pxb,Izubuchi:2018srq}, 
\begin{eqnarray}
    H(\lambda, z^2) &= \int^\lambda_0 C(\alpha_s(\mu),\lambda/\lambda',(z\mu)^2) h(\lambda',\mu^2)d\lambda' \nonumber \\ &+ z^2 C_4(\alpha_s(\mu),\lambda,(z\mu)^2)\otimes h_4(\lambda,\mu^2) + ...
    \label{expansion}
\end{eqnarray}
where the second term
on the right-hand side is a power-suppressed twist-4 contribution. $C$ and $C_4$ are perturbation series in $\alpha_s$. 
For a reasonably large but perturbative
$z^2$, say (0.2~fm)$^2$, judging from the Wilson coefficient function of the number operator $(n=0)$ shown in Fig. \ref{fig:wilson}, where the higher-twist contribution $C_4\otimes h_4$ (the circled product indicates a convolution) may be dropped, one can extract the twist-2 correlation $h(\lambda, \mu^2)$
up to a range about $\lambda_{\rm max} = zP^z\sim 3$ for $P^z=3$ GeV.

Thus according to Eq. (\ref{eq:expand}), instead of calculating the lowest few moments $M_n$, one can equivalently calculate the twist-2 correlation $h(\lambda,\mu^2 )$ in $[0,\lambda_{\rm max}]$ from the coordinate-space correlators.
There exists no accurate estimate of the relation between $\lambda_{\rm max}$ and the maximal order of moment $n$ to be taken into account in Eq.~(\ref{eq:expand}), but these two quantities must be roughly proportional, i.e., the larger $\lambda_{\rm max}$, the more moments one can extract from the correlations~\cite{Bali:2017gfr,Bali:2018spj,RQCD:2019osh, Joo:2019bzr,Joo:2020spy,Gao:2020ito,Detmold:2021qln}.

In Ref. \cite{Gao:2020ito}, the first few moments, $\langle x^2\rangle $, $\langle x^4\rangle $, and $\langle x^6\rangle $,
of the pion valence PDF have been extracted from fitting to the quasi-PDF correlation in the range of $z\in [0,z_{\rm max}\sim 0.8]$~fm. The interpretation of the large $z_{\rm max}$ data in OPE 
is only possible with fixed-order perturbative coefficient functions. The
resummation of large logarithms in $\mu^2z^2$ modify the coefficient
functions drastically at large $z$, as shown in Fig. \ref{fig:wilson}. Thus in an improved analysis with next-to-leading order (NLO) resummation~\cite{Gao:2021hxl}, only data up to $z_{\rm max} \sim 0.5$
~fm have been used, and the determination of $\langle x^6\rangle $ was impossible. Moreover, if one takes into account the uncertainty in the scale setting in the resummed coefficient functions at $z> 0.2$~fm, particularly if introducing the resummation at the next-to-next-to leading order, the error on $\langle x^4\rangle$ will become very larger~\cite{Su:2022fiu}. In addition, non-perturbative contributions may contaminate the fitting for $z> 0.2$~fm, making the moment calculation with local operators an interesting alternative~\cite{Alexandrou:2021mmi}.

Knowledge of twist-2 correlations in a finite range yields global information about PFDs, which, however, is  insufficient to accurately constrain their $x$-dependence. This is similar to the quantum 
mechanical uncertainty principle: Only information in an infinite range in coordinate space provides precise 
momentum space properties. The correlations in $[0,\lambda_{\rm max}]$ allow a resolution of $\Delta x\sim 1/\lambda_{\rm max}$ in momentum space. $\lambda_{\rm max}=3$ corresponds to a range of about 0.3 in $\Delta x$. The SD-OPE constrains the average behavior of PDFs in this range of $\Delta x$. 

Thus any sharp construction of $f(x)$ from $g(\lambda)$ in currently accessible $[0,\lambda_{\rm max}]$ necessarily involves additional assumptions about $f(x)$ in any inverse-problem method~\cite{Karpie:2019eiq}, implicitly modelling the coordinate-space information beyond $\lambda_{\rm max}$. Using a model for $f(x)$ with few parameters is a way to
correlate large and small $\lambda$, and one can fit  $g(\lambda)$ in $\lambda \in [0,\lambda_{\rm max}]$ to 
fix the full $f(x)$. This can be viewed as a generalization to the moments-fit~\cite{Detmold:2003tm,Alexandrou:2021mmi}. The key point, however, is that these fits will have model dependence which is hard to get rid of, unless one has either a large number of moments or a large $\lambda_{\rm max}$, as in global PDF analyses with a large number of data points~\cite{Gao:2017yyd}. 

\subsection{How short is short?}

To reduce the model uncertainty in global fits, one can try to use as large $\lambda$ data as possible from lattice. With the current $P^z$, this means to use data at $z\gg 0.2-0.3$~fm. We found that in the literature, data even up to 0.8~fm or more have been used to make SD-OPE analysis~\cite{Karpie:2019eiq,Sufian:2019bol,Joo:2019jct,Joo:2019bzr,Sufian:2020vzb,Bhat:2020ktg,Gao:2020ito,DelDebbio:2020rgv,Joo:2020spy,Karpie:2021pap}. 
These analysis unfortunately will introduce contamination from non-perturbative contributions which are hard to estimate and subtract (one of those is due to the ratio method of renormalization at large $z$). The important question is how small $z$ must be such that one can justifiably uses SD-OPE to interpret the data.

Short-distance OPE is a double expansion, in terms of the strong coupling constant $\alpha_s$
and the power corrections $z\Lambda_{\rm QCD}$.
Both expansions are related. Let's consider first the $\alpha_s$-expansion in $\overline{\rm MS}$ scheme mostly used in the literature. 

The running of $\alpha_s(\mu)$ is defined
perturbatively only for a limited
range of momenta. The next-to-leading order beta function is definitely no longer perturbative at around $\mu\sim 0.6$ GeV, where $\alpha_s\sim 0.9$. Therefore, the perturbative $\overline{\rm MS}$ momentum scale cannot be less than about $0.6\sim 0.7$ GeV, taking into account the effects of higher-order running. Naively, this corresponds to a distance scale $z\sim 1/\mu\sim 0.35~{\rm fm}$.

However, the scale $\mu$ in any perturbation series can be chosen arbitrarily. Different choices lead to different coefficients in the expansion, which compensate perfectly when a series is known to infinite order. Different choices, however, will lead to different speed of convergences. When a perturbation series is computed only to a finite order, one naturally seeks an optimal scale such that any bigger and smaller one will lead to slower convergence. This is usually dictated by the physics scales under consideration. If there is one physics scale, such as the short distance $z$, perturbation series will contain logarithms of
type $\ln^n(z\mu)^2$ (there is usually a factor of $e^{\gamma_E}/2$ associated with $z$, which is $\sim$ 1), which can become large if $z\mu$ strongly deviates from 1. Thus the most natural choice of the scale is $\mu\sim \delta/z$, where $\delta$ is a constant of order 1. The general practice in the perturbative QCD community is to choose $\delta =1/2\sim 2$ as an estimation of error in perturbation theory. 

The second expansion in OPE is in powers of the small parameter $(z\Lambda_{\rm QCD})^2$. Usually, these contributions are smaller than the leading perturbative terms when $z$ is small. However, 
when the perturbative expansion becomes problematic, power corrections also 
run out of control, 
which happens also around $\alpha_s\sim 1$. The power corrections are strongly  influenced by the non-perturbative QCD vacuum. The short-distance expansion relies on the assumption that the propagation of quarks and gluons is
given by plane waves and their perturbative scattering. However, at larger distance $z\sim 1/\Lambda_{\rm QCD}$, the vacuum starts to distort the plane-wave propagation strongly. 
Thus an alternative criterion for the breakdown of SD-OPE is that  non-perturbative vacuum effects become important. 

At a simple level, the QCD vacuum properties are
described by various condensates. The most basic 
is the gluon condensate $\langle (\alpha_s/\pi)G^2\rangle$, which has been used in QCD sum rule calculations~\cite{Shifman:1978bx}.
A recent first-principles determination of the condensate is $(1.33r_0^{-1}\rm GeV)^4 = (0.53 \rm GeV)^4$~\cite{Ayala:2020pxq}. This scale indicates a breakdown of the twist expansion at distance scale $0.75r_0\sim 0.4~{\rm fm}$ where the power terms have the size of the leading term, which is consistent with the estimation from the running coupling above. A well studied model for the QCD vacuum is the instanton liquid~\cite{Schafer:1996wv}, in which  the classical gauge field configurations consisting of various sizes of instantons generating the non-perturbative physics. One of the most important scales is the average instanton size, 
about 0.3~fm. Thus the gluonic instanton configurations will have strong influences on any quark and gluon propagation at distance scale larger than 0.3~fm. 

\begin{figure}[tbh]
\centering
\includegraphics[width=0.42\textwidth]{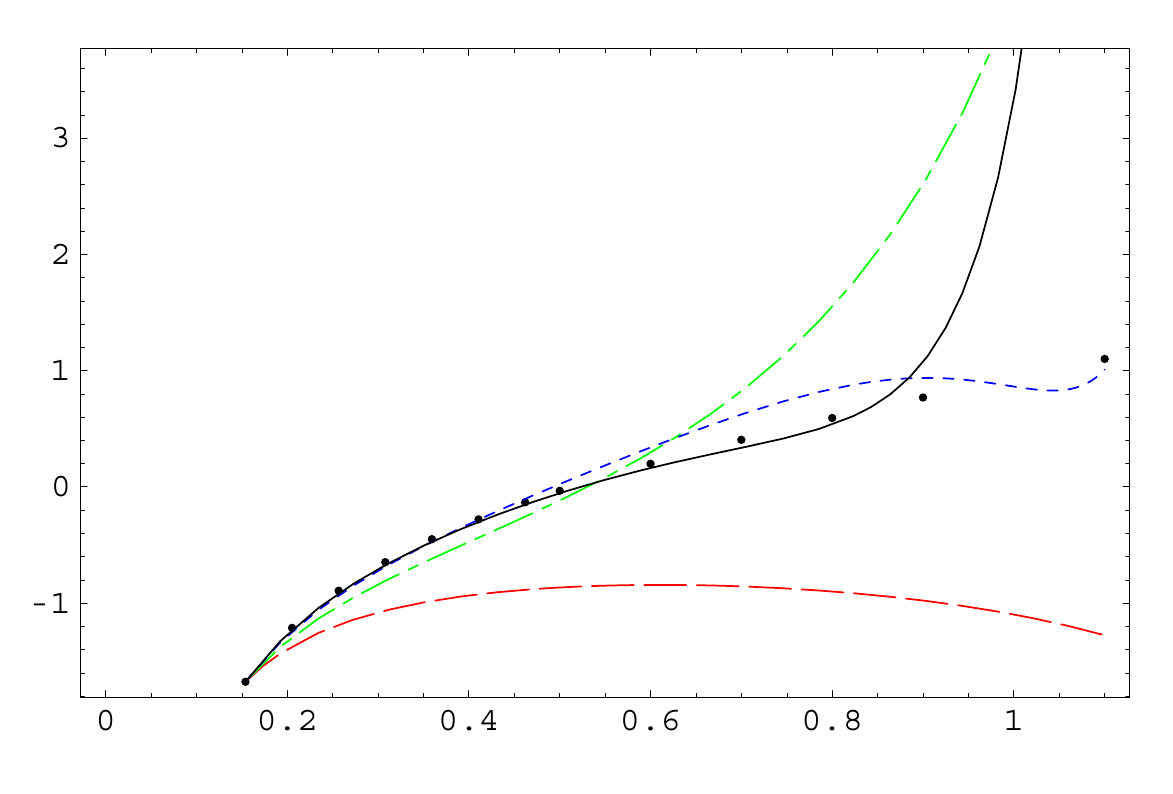}
\centering
\caption{Heavy-quark potential versus $r$ at tree (red dashed line), one-loop
(green dash-dotted line), two-loop (blue dotted line) and estimated three-loop plus the RG improvement 
for the ultrasoft logs (solid line) compared with the lattice results (dots) from~\cite{Pineda:2002se}. The horizontal distance scale is in units of $r_0\sim 0.5$~fm.}
    \label{fig:pot}
\end{figure}

\begin{figure}[tbh]
\centering
\includegraphics[width=0.5\textwidth]{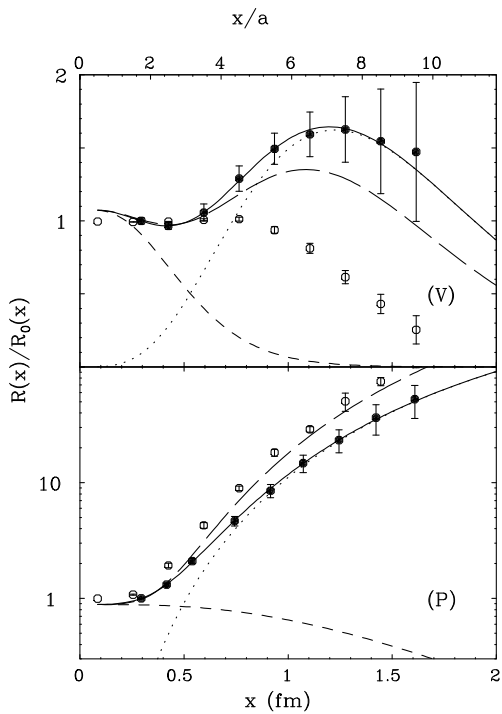}
\centering
\caption{Vector and pseudo-scalar density correlation functions. The short dashed line is the perturbative contributions, and the dots are resonace contribution. Solids dots are
from lattice calculations~\cite{Chu:1993cn}.}
    \label{fig:density}
\end{figure}

\begin{figure}[tbh]
\centering
\includegraphics[width=0.41\textwidth]{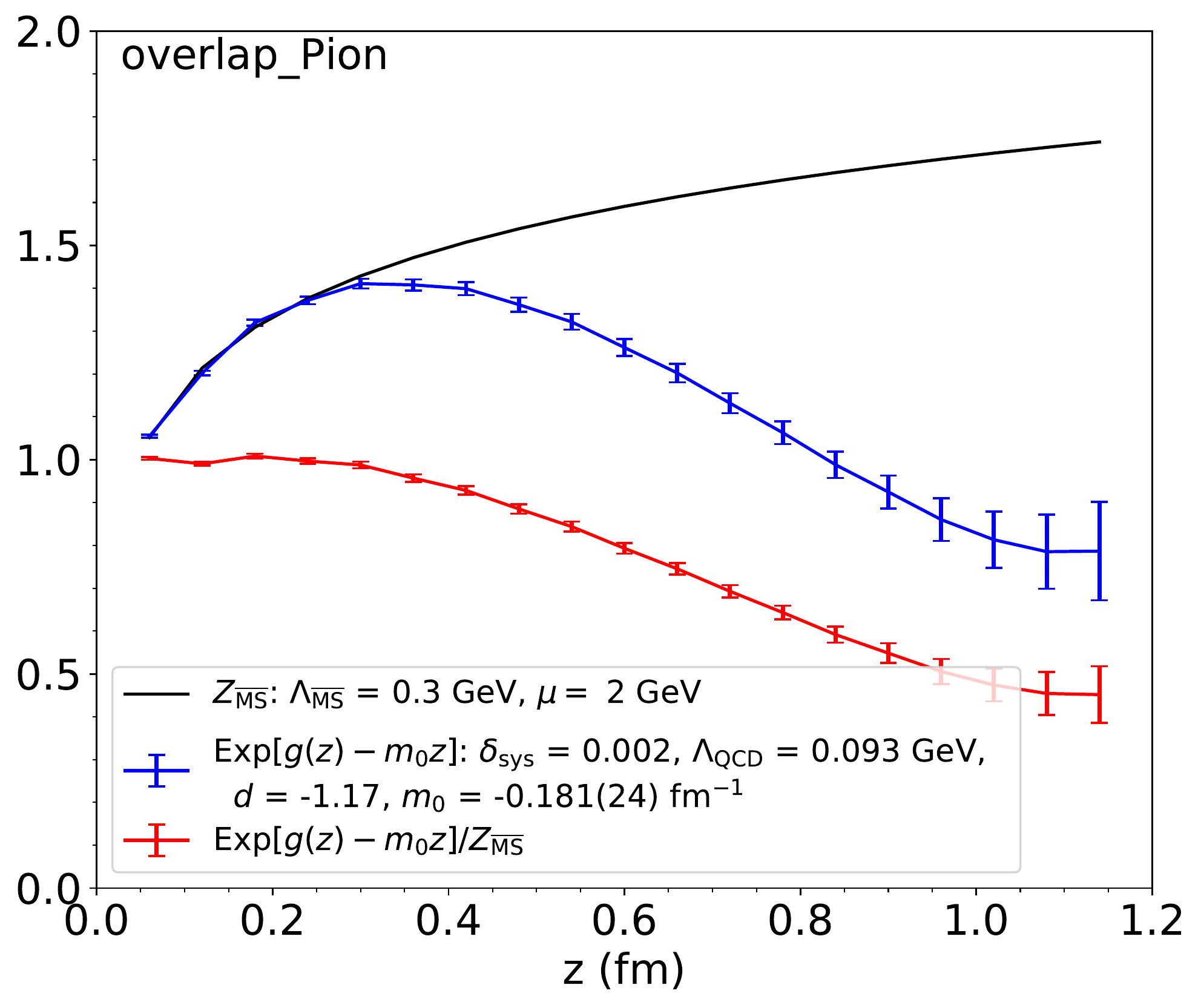}
\centering
\caption{The renormalized quasi-PDF correlator in the zero-momenutm pion state (blue dost), compared with NLO perturbation theory calculations. The red dots are the ratio of the two~~\cite{LatticePartonCollaborationLPC:2021xdx}.}
\label{fig:qPDF}
\end{figure}

\begin{figure}[tbh]
\centering
\includegraphics[width=0.47\textwidth]{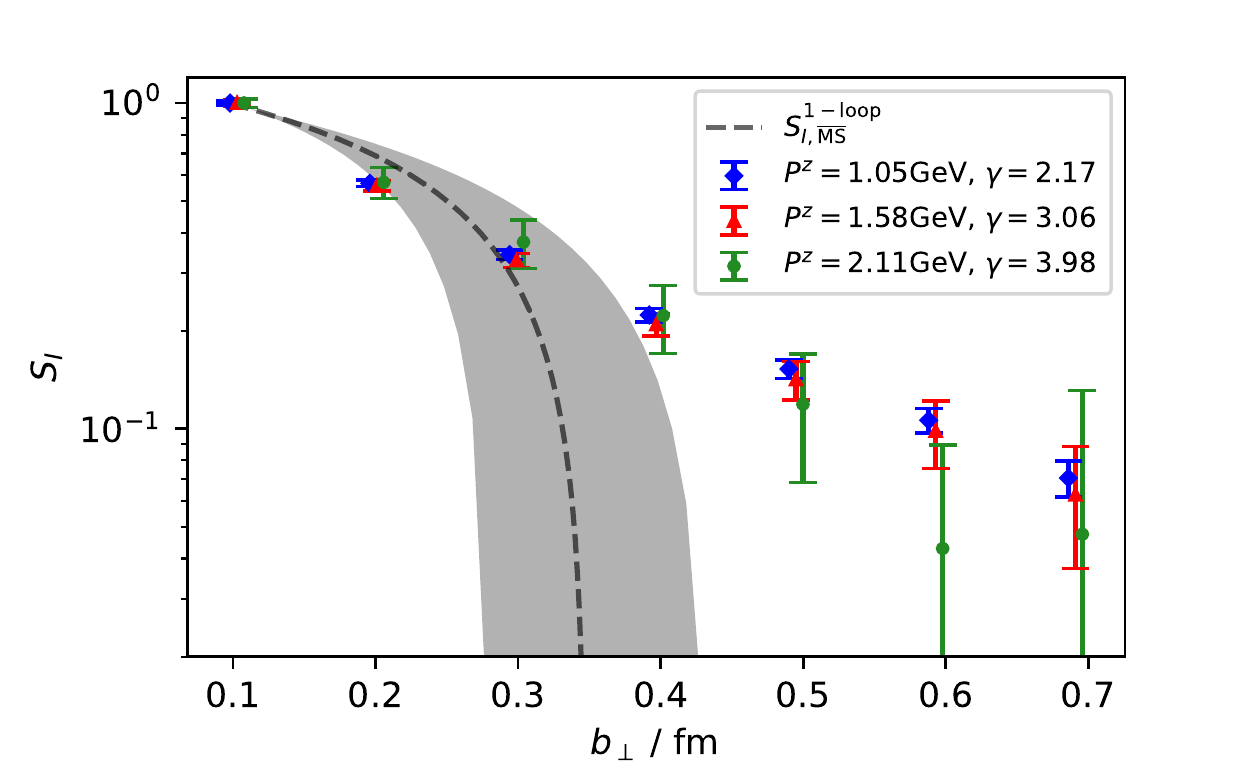}
\centering
\caption{Soft function in TMD factorization. Data are from lattice, and dashed line from perturbation theory~\cite{LatticeParton:2020uhz}. }
    \label{fig:sfunction}
\end{figure}

The non-perturbative vacuum effects on the quark and gluon propagators can be studied on lattice in a fixed gauge~\cite{Huber:2018ned}. In the so-called maximally Abelian gauge where the Abelian gluons dominate the non-perturbative physics, the color confinement mechanism resembles a dual superconductor~\cite{Nambu:1974zg,Mandelstam:1974pi,tHooft:1977nqb}. In recent lattice studies~\cite{Gongyo:2013sha, Schrock:2015pna}, it was found that the Abelian (diagonal) gluons acquire an effective mass of order 0.36 GeV, which indicates that the non-perturbative effect on the gluon propagation reaches the level of 50\% at the distance scale at about 0.4~fm. 

The above discussion is completely general, independent of the spatial correlators under consideration. Both QCD perturbation theory and twist-expansion hit a hard wall at around $z=0.3\sim 0.4~$fm, beyond which physics become entirely non-perturbative. To stay in the perturbative region where higher-twist contributions are at the level of $10\%$ or less, the distance scale must be smaller than 0.2~fm. In fact, if QCD physics became entirely perturbative at $0.2$~fm, lattice QCD with perfect-action simulations at this lattice spacing would have produced satisfactory results, which is not the case~\cite{DeGrand:1994zr}. 

A number of case studies on spatial correlations confirm the above consideration:
\begin{itemize}
    \item Potential between static sources~\cite{Pineda:2002se,Bazavov:2014soa,Ayala:2020odx}. The studies show that perturbation theory works fine up to $0.25$~fm, breaks down above $0.4$~fm where non-perturbative effects set in quickly, as shown in Fig. \ref{fig:pot}.
    
    \item Hadron-hadron correlators ~\cite{Chu:1993cn}. Lattice QCD simulations have shown that the
perturbative and non-perturbative contributions become about equal at around $z=0.5$~fm. Perturbative results
are fairly accurate until 0.2~fm, as shown in Fig. \ref{fig:density}. 

\item Quasi-PDF correlators in zero-momentum hadron states~\cite{LatticePartonCollaborationLPC:2021xdx}. After
the careful ultraviolet renormalization, the quasi-PDF correlators are found to match to NLO perturbation theory well up to 0.25~fm, beyond which deviations occur as shown in Fig. \ref{fig:qPDF}. 

    \item QCD soft function in TMD factorization~\cite{Collins:1981va}, which is a function of transverse distance $b$: For $b$ larger than $0.3$~fm, the perturbative expansion breaks down~\cite{LatticeParton:2020uhz}, as shown in Fig. \ref{fig:sfunction}. 
\end{itemize}

Thus the perturbation series $C$ and twist expansion in Eq.(\ref{expansion}) break down at large $z\sim 0.4$~fm.  Moreover, for $z>0.2$~fm the correlation in $G(\lambda)$ is contaminated with non-perturbative contributions, resulting in possibly large bias of the moment analyses and global PDF fits using inverse-problem methods, which are difficult to access quantitatively~\cite{Gao:2020ito,Joo:2020spy,Karpie:2021pap}.

\section{Direct Calculations
of PDFs in $x$-space using LaMET}

Large-momentum effective theory aims at calculating parton densities directly at a 
fixed $x$, which we refer to as {\it local information} of PDFs, when the latter falls into the range where the large-momentum expansion converges.

According to Feynman, partons emerge from the infinite momentum limit of {\it momentum densities} in hadrons. LaMET expansion begins from the quark or gluon momentum distributions $N(k^z=xP^z,P^z)$  at a large but finite hadron momentum $P^z$ in $z$-direction, where $k^z$ is the longitudinal momentum of a constituent with trasverse momentum $\vec{k}_\perp$ integrated out, and takes it as a zero-order approximation (called quasi-PDF in the literature). This is not unreasonable because the difference in momentum distributions between a large and the infinite momentum can naively be assessed by a dimensionless parameter $\Lambda_{\rm QCD}/P^z$. If $P^z\gg \Lambda_{\rm QCD}$, the hadron structure at this $P^z$ may considered as being in the asymptotic region, where the momentum distribution shall be similar to the PDFs at infinite momentum. This can be viewed as {\it large-momentum symmetry} in analogy to heavy quark symmetry~\cite{Georgi:1990um}, in the sense that hadrons travelling at 10 GeV in energy have momentum distributions similar to those travelling at 10 TeV. In field theories, however, there are large logarithms in $P^z$, Feynman's idea of momentum-independence can only be realized through effective field theory~\cite{Ji:2014gla}.

Calculating the momentum density $N(k^z, P^z)$
for gauge theories is a bit involved. One needs to start from the coordinate 
space correlation
\begin{equation}
   h(z, P^z)= \langle P^z| \psi^\dagger(z)\psi(0)|P^z\rangle \ , 
\end{equation}
at all $-\infty<z<\infty$, which can be  difficult for large $z$ where the numerical signals become noisy on a lattice. Moreover, quarks and gluons are colored particles, and gauge symmetry requires connecting the field operators either by a colored propagator or a Wilson line $W(z)$. The straight Wilson line is a preferred choice
for computational simplicity. However, it leads to linear divergent self-energy in lattice spacing $a$~\cite{Ji:2017oey,Green:2017xeu,Ishikawa:2017faj}, which must be renormalized precisely. 

The linear divergences in coordinate-space correlations $h(z)$ can be subtracted using the standard mass renormalization~\cite{Ishikawa:2016znu,Chen:2016fxx,LatticePartonCollaborationLPC:2021xdx}. 
To reduce the discretization error near $z\sim 0$, where $z\to 0$ does not commute with $a\to 0$, 
a hybrid scheme has been introduced in which the short-distance correlation function at $z<z_S\sim 0.2$~fm is renormalized with a $z$-dependent lattice matrix element $Z_o(z,a)$, 
\begin{equation}
    h^R(\lambda,P^z/\mu_X) = \left(Z_o(z,\mu_X)/Z_o(z,a)\right) h(z,a)  \ , 
\end{equation}
and $Z_o(z,\mu_X)$ is the same matrix element calculated in $X$-scheme, e.g., in $X=\overline{\rm MS}$. 
For $z>z_S$, 
\begin{equation}
    h^R(\lambda,P^z/\mu_X) = C(\mu_X,a)e^{\delta m(a)z} h(z,a)
\end{equation}
where $C(\mu_X,a)$ is related perturbatively to the anomalous dimension of the heavy-light current~\cite{Ji:1991pr}, and can be determined non-perturbatively by matching to the short-distance $h^R$ at the boundary $z=z_S$. 

On the other hand, it is well-known that the pole mass has the so-called infrared renormalon problem in the sense that the perturbative series for the linear-divergent mass does not converge~\cite{tHooft:1977xjm}. Therefore, it is important to check the renormalon consistency of the hybrid renormalization scheme. Indeed, the perturbation series $Z_o(z,\mu)$ has the same renormalon uncertainty as that in $\delta m(a)$~\cite{Braun:2018brg}, and therefore
renormalons in both regions can be matched at $z=z_S$. 

One can, without loss of generality, use the principal-value (PV) prescription
to define the perturbation series for both $Z_o(z,\mu)$ and $\delta m(a)$~\cite{Ayala:2019uaw}. After subtracting the linearly-divergent mass counter term, the correlation function decays exponentially 
at large $z$~\cite{Burkardt:1994pw}
\begin{equation}
    h^R(z,\mu) \sim \exp(-\bar \Lambda z) \ , 
\end{equation}
where $\bar \Lambda $ is the binding energy of a light-quark to a color source when the mass of the color source is calculated with a PV prescription. Studies have show that $\bar \Lambda $ is on the order of few MeV or so, which corresponds to the size of a typical hadron~\cite{Ayala:2019hkn}. This asymptotic behavior is crucial for lattice calculations as the signal-to-noise ratio decays quickly at large $z$, and the missing information can impose significant uncertainty of Fourier transformation. Thus the exponential decay mainly serves to reduce the uncertainties in the large-$z$ correlation, and one can safely calculate the correlation functions up to the region around 1$\sim $1.5 fm to permit a physical extrapolation beyond this. The use of this physical extrapolation in LaMET is to reduce the uncertainties in the momentum density calculations of the middle-$x$ region~\cite{Gao:2021dbh}, not about predictions of PDFs at small-$x$ where the expansion breaks down.

Thus, from the lattice data, one can obtain non-perturbative coordinate-space correlations in the  $\overline{\rm MS}$ scheme. 
Due to the renormalization scale dependence $h^{\overline {\rm MS}}(\lambda, P^z/\mu)$ has a discontinuity at $\lambda=0$. This can be improved using perturbation theory~\cite{Izubuchi:2018srq}. The renormalized momentum distribution is just a Fourier transformation,  
\begin{equation}
N^R(y,P^z/\mu) = \int^\infty_{-\infty}
d\lambda e^{i\lambda y}
h^R(\lambda, P^z/\mu)
\end{equation}
where $y$ has support $-\infty<y<\infty$, which is physically sensible for a hadron with finite momentum. Using $N^R(x,P^z)$, one can make  a large-momentum expansion for PDFs,
\begin{eqnarray}
   f(x,\mu^2) &=& C_N(x,P^z/\mu)\otimes N^R(x, P^z/\mu) \nonumber \\
        &+& C_4(x, P^z/\mu)\otimes f_{4}^R(x, P^z) \left(\frac{M}{P^z}\right)^2
        + ... \ , 
\end{eqnarray}
where $C_N$ and $C_4$ are perturbation series in $\alpha_s(P^z)$, and take into account the difference between the large momentum limits with renormalization done before hand, e.g., physical $N^R$ in which there is a large logarithms of $P^z$, and with the renormalization done afterward which defines the standard PDFs with partons as effective theory object~\cite{Ji:2014gla}. $C_N$ contains the renormalon uncertainty corresponding to the mass subtraction \cite{Braun:2018brg} which must be regulated likewise by a PV prescription. The key point of the above expression is that the infrared physics of momentum distributions in the large $P^z$ limit is the same as the light-cone distribution $f(x,\mu^2)$: switching the limit $P^z\to\infty$ and $a\to 0$ 
does not change the soft and collinear infrared physics~\cite{Ji:2014gla}. 

Power counting is a key to any effective field theory~\cite{Weinberg:2021exr}.
Here the small parameter is the bound state scale $\Lambda_{\rm QCD}$ or hadron masses. 
The obvious high-energy scale is the momentum $P^z$.
However, a careful examination indicates the momentum of the active particle $k^z$ and the remnant momentum $P^z-k^z$ can also act as
high-energy scales. Thus a conservative systematic power counting is in powers of $(\Lambda_{\rm QCD}/k^z)^2 \sim 1/(xP^z)^2$
and $(\Lambda_{\rm QCD}/(P^z-k^z))^2\sim 1/((1-x)P^z)^2$.
Therefore, LaMET can only produce 
result in a region of $x\in [x_{\rm min}\sim \Lambda_{\rm QCD}/(xP^z), x_{\rm max}\sim 1-x_{\rm min}]$ where the higher-order power contributions
are small. The expansion is expected to break done very quickly near the end-point regions.

\subsection{How large is large?}

Large-momentum effective theory requires large hadron momentum to work $P^z\gg \Lambda_{\rm QCD}$. Similar to the discussion for short-distance factorization, how large a momentum is large in LaMET? The power counting gives a more accurate answer: $xP^z$ and $1-x)P^z$ must be large perturbative scales, for example, $~1$ GeV. 

A more refined result comes from the perturbative QCD matching and renormalon analysis. The one-loop matching coefficient contains the following logarithm~\cite{Izubuchi:2018srq}
\begin{equation}
    \ln (\mu^2/4x^2(P^z)^2)
\end{equation}
which conjugates to $\ln \mu^2z^2$ in coordinate space. Thus it appears that $\mu = 2xP_z$ is the right scale setting for higher-order resummation, which correspoinds to $xP^z\sim 0.4$~GeV. With
$P^z$ around 2.4 GeV, the fixed-order perturbative matching shall be accurate down to $x\sim 0.15-0.2$~\cite{Su:2022fiu}.

Note that there are soft-radiation dominating terms in the matching kernel which goes like $\ln(1-\xi)/(1-\xi)$. For $x$ in the middle region, these term do not generate large contributions due to cancellation from $\xi>1$ and $\xi<1$. However, 
near the end point $x=1$, the cancellation is incomplete, and large threshold logarithms remain. These large threshold logs
will effectively lower $P^z$ and generates an effective scale $4(1-x)(P^z)^2$. This conclusion
is similar to the renormalon calculation in \cite{Braun:2018brg}. Thus the large-momentum expansion can potentially reach closer to $x=1$ than $x=0$. This is very interesting because phenomenological PDFs at large-$x$ currently have large uncertainties.  However, this enhancement of accuracy from the above analysis has not been observed in LaMET calculations so far, which indicates that there might be other systematics that must be corrected  before a reliable large-$x$ prediction can be made. 

\section{Short-Distance-Expansion-Assisted PDFs in $x$-space}

The relationship between SD-OPE and LaMET is simple in the infinite momentum limit: they are two equivalent ways to define partons. However, they lead to two different approximation schemes for analyzing real-world data 
at finite hadron momenta. They are not identical expansions and will not obtain the same result from the same data, as one might have guessed naively. So which one is a better choice for which type of problem? To which degree are both approaches complementary? We try to discuss these questions in this section. 

\subsection{Contrast}

If one's ultimate goal is to obtain twist-2 coordinate-space correlations, corresponding to the global properties of partons, SD-OPE is the right choice, resulting in a segment $[0,\lambda_{\rm max}]$ of these from the finite-momentum lattice data. On the other hand, if one is interested in $x$-space PDFs, LaMET is the the natural method to get a range $[x_{\rm min}, x_{\rm max}]$ of distributions with  controlled precision.
The segments of the functions in separate coordinate and momentum spaces cannot be naively translated into each other through Fourier transformation, except in the infinite momentum limit. 

One important reason that the LaMET approach
allows to obtain more precise information for $x$-space PDFs is that the momentum-space expansion utilizes the full $z$-range of 
data for the coordinate-space correlations, and filter out their higher-twist contributions through Fourier transformation. No manual twist separation is necessary for the coordinate-space correlations even though the twist expansion itself breaks down beyond $z\sim 0.4$~fm~\cite{Ji:2020brr}.  Therefore, LaMET calculations use {\it legitimately} more lattice data than SD-OPE would allow. The correlation data at large-$z$ are crucially important up until the exponential decay region, and the full-range correlations ensure that after Fourier transformation, the momentum distributions are of pure twist-2 in the region $x\in [x_{\rm min}, x_{\rm max}]$. 

On the other hand, SD-OPE can only use the correlation data up to $z\sim 0.2$~fm, beyond which one has to subtract the non-perturbative contributions, an exercise hard to control systematically. Indeed, it has been a great challenge in perturbative QCD to understand higher-twist effects {\it quantitatively} because they are intimately related to higher-order perturbation theory at the leading twist. The leading higher-twist contributions have been studied from first principles only in few cases: heavy-quark potential~\cite{Pineda:2002se,Ayala:2020odx}, heavy quark masses~\cite{Ayala:2019hkn}, and QCD vacuum gluon condensate~\cite{Ayala:2020pxq}. Moreover, as said repeatedly, the twist expansion breaks down at $z\sim 0.3-0.4$~fm beyond which the perturbative QCD interpretation of data is impossible. 

If just using the pure twist-2 part of the correlations from SD-OPE, one cannot reconstruct the PDFs at any value of $x$ with controlled accuracy with current lattice data, a disadvantage which cannot be overcome by any inversion methods which miss important physics information at long 
correlation range. A discussion of the comparison between LaMET and SD-OPE analyses was first made in~\cite{Ji:2020byp}, and the results for the pion PFD analyses confirm the main observation of this paper~\cite{Gao:2020ito,Gao:2021dbh}. 

\subsection{Complementarity}

Interestingly, LaMET expansion does not exhaust all constraints on PDFs from the finite-momentum lattice data. SD-OPE analyses can provide complementary information which can help LaMET analyses. An important application of SD-OPE is to study the $aP^z$ artefacts of lattice data~\cite{Joo:2020spy, Gao:2020ito}. Here we
focus on the global constraints on parton
physics from twist-2 correlations. 
Coupled with theoretical and phenomenological arguments that PDFs behave like $x^\alpha$ at small-$x$ (Regge analysis \cite{Ioffe:1985ep}) and $(1-x)^\beta$ at large-$x$~\cite{Drell:1969km, West:1970av}, short-distance correlations can be
used to improve upon LaMET calculations in the end-point regions $[0, x_{\rm min}]$ and $[x_{\rm max},1]$. 

It is straightforward is to parameterize $f(x)$ in these regions in terms of a systematic expansion,
\begin{eqnarray}
     f(x) & =&  Ax^\alpha (1+a\sqrt{x}+...),  ~~~~~~~~~~~~~x\in [0, x_{\rm min}]  \\
        &=& B (1-x)^\beta \Big(1+b(1-x)+...\Big), x \in [x_{\rm max}, 1] \nonumber \ ,  
\end{eqnarray}
where we have included the next-to-leading terms at small and large $x$. The inclusion
of $\sqrt{x}$ is purely phenomenological and one could try analytic terms in $x$ as well~\cite{Dulat:2015mca}. 
One can also use more sophisticated parametrizations if well motivated or even neural network approaches as in the global analyses. The normalization $A$ and $B$ can be determined by continuity conditions at $x=x_{\rm min}, x_{\rm max}$. More continuity conditions such as first-order derivatives can be added as
additional constraints. All parameters including the exponents $\alpha$ and $\beta$ can be fitted to either twist-2 
correlations or lowest order moments, with fixed LaMET predictions on PDFs in the central-$x$ region. This presumably will generate state-of-art lattice-QCD 
PDFs with minimal bias. 

\begin{figure}[tbh]
\centering
\includegraphics[width=0.47\textwidth]{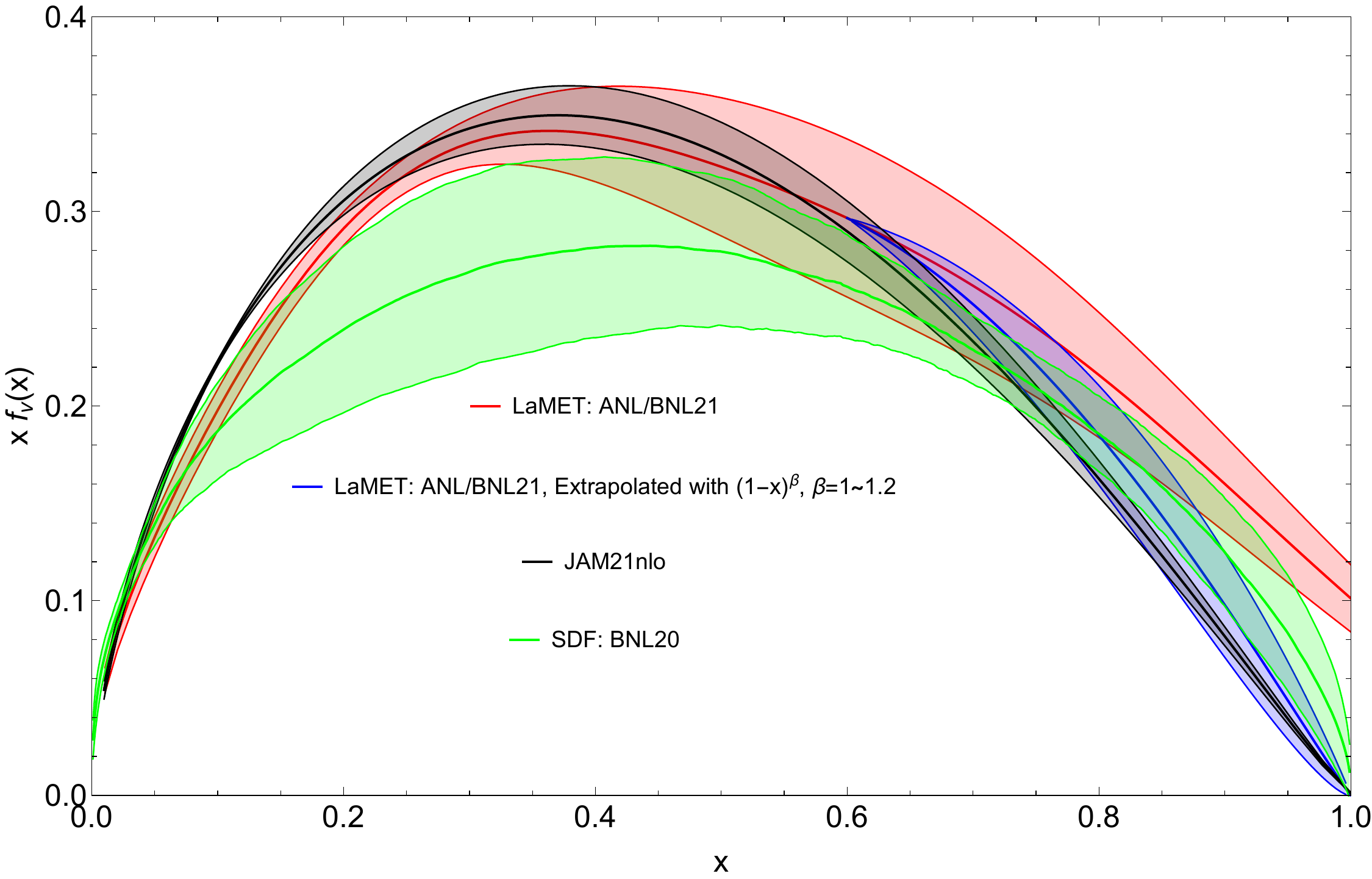}
\centering
\caption{Valence pion PDFs from both LaMET (red band) and SD-OPE (green band). The blue band is a fit to the second moment of the PDF plus LaMET result up to $x=0.6$. The black band is the JAMnlo result from fitting to
experimental data.}
    \label{fig:pion}
\end{figure}

As an example, let us consider
the pion valence distribution. The ANL/BNL group has recently produced
some of the most accurate data for coordinate space correlations in terms of quasi-PDFs~\cite{Gao:2020ito, Gao:2021dbh}. The data was generated for pion mass 
$m_\pi \sim 300$ MeV at very small 
lattice spacings $a=0.06$ fm and 0.04 fm. The data has been analyzed in both short-distance
expansion~\cite{Gao:2020ito} and LaMET~\cite{Gao:2021dbh}, with results
shown as red (LaMET) and green (SD-OPE) bands in Fig. \ref{fig:pion}.  

As one can see, the LaMET calculation generates a reasonable prediction for the valence PDF in the $x$-range roughly from 0.1 to 0.7.
For $x<0.1$ and $x>0.7$, LaMET results
become less reliable due to higher-twists, and ultimately the momentum-space expansion breaks down.

On the other hand, the $x$-dependent
analysis of the short-distance expansion is model-dependent~\cite{Gao:2020ito}. In particular in the middle region, the result has a large error. This unreliability propagates to the larger-$x$ region because of the global constraint from the short-distance correlation or lower moments. This results in a very stiff large-$x$ distribution~\cite{Gao:2020ito}.

If one uses the LaMET result
in the middle $x$ region and fits the 
phenomenological parametrizations
using the global constraints from 
the small-distance expansion, or the first
few moments, 
the distribution at large-$x$ will 
improve. As a quick exercise, we use $A(1-x)^\beta$ beyond $x_{\rm max}=0.6$ as a fitting function, and constrain  $\beta$ by the second moment $\langle x^2\rangle$ which has been calculated to high accuracy in both Refs.~\cite{Gao:2020ito,Gao:2021dbh} and~\cite{Alexandrou:2020gxs}. This simple fit yields an exponent $\beta = 1.0\sim 1.2$,
corresponding $\langle x^2\rangle= 0.110$ to 0.104, shown as blue band in Fig. \ref{fig:pion}. This large-$x$ result is more consistent with JAM fit than the SD-OPE fit. If one chooses $x_{\rm max}= 0.7$ the large-$x$ behavior will be further  softened. 
A full fitting analysis to the entire twist-2 coordinate-space correlation including both small and large $x$ regions is beyond the scope of this
article. 

Finally, one might wonder how it is possible that the same data analyzed differently can generate complementary information on the PDFs. For example, why does the LaMET analysis not already exhaust all useful information? An 
answer follows from the first moment of PDFs. LaMET calculations are
not trustable in the end-point regions and therefore cannot predict reliably  the first moments. On the other hand, one can use the parton phenomenology and global properties to fix PDFs in these regions.  However, for a large $P^z$, the end-point regions are small, and the contributions from these regions to lower moments are also small. If there is significant uncertainty in the lower moments, coupled with errors in the LaMET analysis, the
constraints for the end-point regions 
could be considerably weakened. Therefore, the above strategy may only be useful at moderate $P^z$.

\section{Conclusion}

In this article, We have explained the similarities and differences between SD-OPE and LaMET. In particular, we have shown how to use the twist-2 correlation functions from SD-OPE to phenomenologically improve the LaMET predictions of PDFs in the end-point regions. 
We have used the recent result on the pion valence PDF as an example to demonstrate this point. Immediate applications can be made to the proton's iso-vector PDF
and meson distribution amplitudes~\cite{Hua:2022kcm}. Furthermore, the approach discussed is entirely applicable to other parton observables, including but not limited to GPDs and TMD PDFs.  

\section*{Acknowledgments}

I thank G. Bali, Peter Petreczky, A. Pineda, Jianhui Zhang and Yong Zhao for many discussions. In particular, I thank A. Schaefer for a very careful reading of the manuscript and Yushan Su for making Fig. 6. This work is partially supported by the U.S. Department of Energy under Contract No. DE-SC0020682.

\bibliographystyle{apsrev4-1}
\bibliography{references}

\end{document}